\documentclass[fleqn,10pt]{wlscirep}
\usepackage[utf8]{inputenc}
\usepackage[T1]{fontenc}
\title{Optical Mode Control, Switching and Shaping In Few Mode Fiber Using a Fiber Piano}

\author[1,*,+]{Shuin Jian Wu}
\author[1,*,+]{Anindya Banerji}
\author[1]{Ankush Sharma}
\author[3]{Zohar Finkelstein}
\author[3]{ Ronen Shekel}
\author[3]{Yaron Bromberg}
\author[1,2,]{Alexander Ling}
\affil[1]{Centre for Quantum Technologies, National University of Singapore, 3 Science Drive 2, Singapore 117543}
\affil[2]{Department of Physics, Faculty of Science, National University of Singapore, 2 Science Drive 3, Singapore 117551}
\affil[3]{Racah Institute of Physics, The Hebrew University of Jerusalem, Jerusalem 91904, Israel}

\affil[*]{cqtwsj@nus.edu.sg, cqtab@nus.edu.sg}

\affil[+]{these authors contributed equally to this work}

\begin{abstract}
This work investigates the use of a fiber piano in controlling spatial modes in few mode fibers. It has been found that together with sub-optimal coupling into SMF-28 fibre and half and quarter waveplates, the fiber piano is capable of producing and reproducing desired spatial modes up to $LP_{11}$ when using 808 nm light and up to $LP_{21}$ when using 632.8 nm light. The control of spatial mode profile extends down to the single photon level. This is demonstrated with the help of correlated photon pairs generated via spontaneous parametric down conversion. 
\end{abstract}
\begin{document}

\flushbottom
\maketitle
%
%
\thispagestyle{empty}


\section*{Introduction}

Although light is a common means of transmitting information in modern communications, not all of its available degrees of freedom (DoF) can be freely controlled \cite{Shen22}. For instance, methods to control polarization and frequency are easily found, but light coupled into a multimode fiber leads to uncontrolled interference between various spatial modes that creates a speckle pattern at the output \cite{saleh19}. Being able to control and sort those spatial modes would allow for higher data transmission rates using multimode fibers as compared to standard single mode fibers \cite{Braverman20}, along with other applications.

In current free space optical communications, including quantum communications \cite{Sidhu}, the signal is often collected into a single mode fiber. This could be quite challenging, especially when the signal passes through turbulent conditions that cause wavefront distortions. In such cases, adaptive optics \cite{Defienne2018} together with active feedback control is necessary to ensure sufficient coupling of the signal into the fiber. Using a multimode or even a few mode fiber \cite{Zheng2016} can relax these requirements a bit due to the larger numerical aperture (NA) of these fibers. This leads to higher coupling efficiencies \cite{Jin} which result in better signal to noise ratios. But, this comes at the cost of the signal getting scrambled due to uncontrolled interference of the spatial modes. Hence, some means of spatial mode control is needed in order to benefit from the higher coupling efficiencies of few or multimode fibers. 

One method of control involves using spatial light modulators to shape the wavefront before it enters the fiber, the goal being to selectively excite only the desired modes to be propagated \cite{Liang2017}. Unfortunately, for a fiber supporting N modes, shaping the input wavefront only affords up to 2N degrees of control, whereas the fiber's transmission matrix contains $N^{2}$ elements \cite{Bromberg2020}. Furthermore, mode groups of order higher than $LP_{01}$ have degenerate modes that easily couple amongst themselves, altering the output spatial profile \cite{Liang2017}. Hence, a more productive approach would be to use the transmission matrix.

With deep learning methods the transmission matrix of a fiber could be solved for in order to compensate for it \cite{Zhang2022}. However, matrix inversion problems are complex and computationally costly, and ultimately a spatial light modulator will be needed to modify the input wavefront to obtain the desired output. Such spatial light modulators tend to be lossy, and could offset the gains from the increased coupling efficiency. In addition, the main means of spatial mode control would still be the spatial light modulator, which as stated before lacks the degree of control available to the transmission matrix \cite{Bromberg2020}.

Rather than simply learn the transmission matrix, it would be better to gain control of it. A fiber piano has been proposed previously as a means to do so. The fiber piano consists of 'keys' made of piezoelectric elements that allow for controlled mechanical perturbations to be performed on a fiber to induce local bends \cite{Bromberg2020,Bromberg2021,Bromberg2024}. These bends allow for coupling across spatial mode groups, allowing for a Particle Swarm Optimization (PSO) algorithm \cite{PSO1,PSO2} to find key configurations that produce the desired spatial modes at the output given a certain input. The fiber piano has been able to perform as a tunable bandpass filter \cite{Bromberg2023_1} as well as increase the coupling from a multimode fiber to a single mode fiber and shape single photons \cite{Bromberg2023_2}. 

In this work, we explore the use of a 15 key fiber piano in guiding single spatial modes through a few mode fiber given an arbitrary superposition of spatial modes excited at the input. We find that spatial mode control can be enhanced with the use of a half waveplate (HWP) and quarter waveplate (QWP) just before light is coupled into the few mode fiber, along with tuning the misalignment of light into said fiber. Once the fiber piano has solved for the desired spatial modes, it can swap back and forth between them easily (see supplementary video). We have managed to obtain $LP_{01}$, $LP_{11}$ and $LP_{21}$ modes using this method, and shown that it works on single photons as well.



\section*{Methods}

\begin{figure}[ht]
\centering
\includegraphics[scale=0.7]{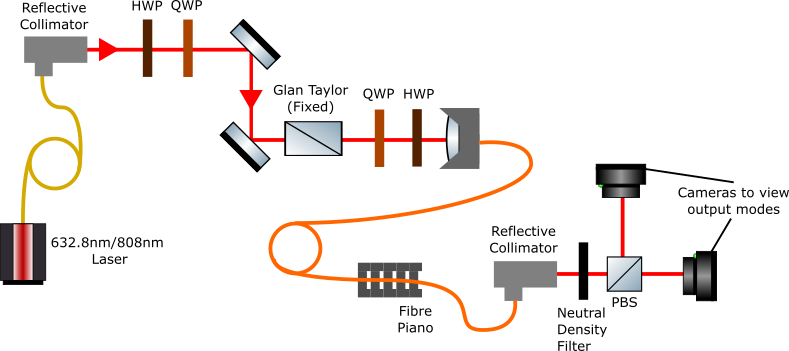}
\caption{Experimental setup for guiding laser light through the few mode fiber with the fiber piano. For testing with single photons the lasers were later swapped with a photon pair source and the camera at the reflected arm of the PBS with an EMCCD.}
\label{fig:lasers}
\end{figure}

\begin{figure}[hbt!]
\centering
\includegraphics[width=\linewidth]{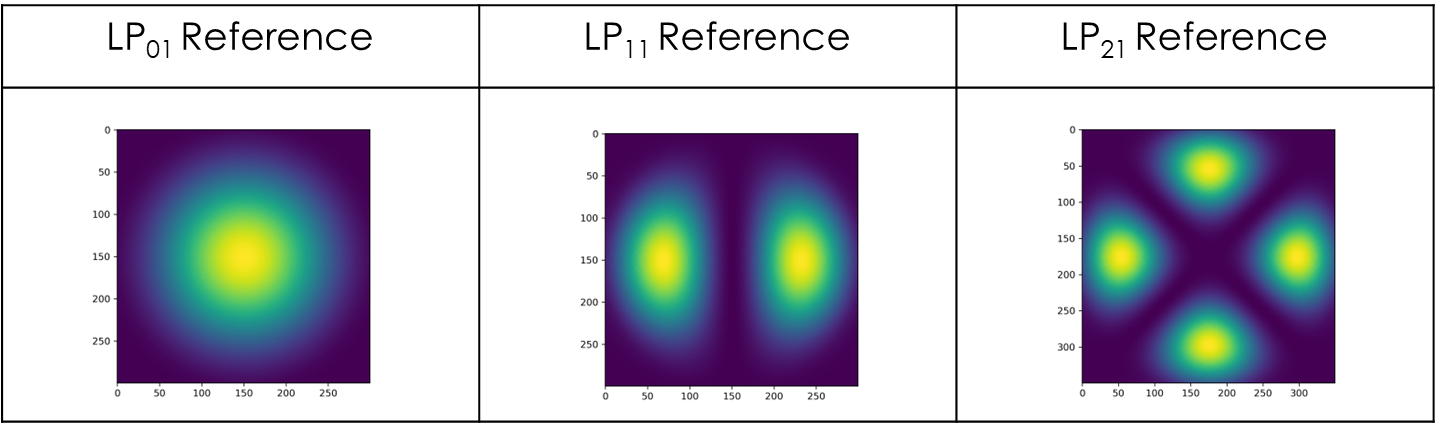}
\caption{Reference images generated from python module ofiber used to calculate 2D correlation (corr2) values with images obtained from the cameras. The $LP_{11}$ and $LP_{21}$ references are rotated at various angles to find the best corr2 values.}
\label{fig:refs}
\end{figure}

The experimental setup used is shown in Figure \ref{fig:lasers}. Light sources were first coupled into 780HP single mode fiber before entering the setup. The 15 'key' (piezo actuator, ThorLabs PB4NB2W) fiber piano used is of the same construction as Bromberg et al's design \cite{Bromberg2020,Bromberg2021,Bromberg2023_1,Bromberg2023_2}. Each actuator creates a three-point contact on the optical fiber running under it. This creates local bends on the fiber of varying radius of curvature when different voltages are applied to the actuators. Testing was done with He-Ne (632.8 nm) and 808 nm lasers, and the few mode fiber used was SMF-28 (NA: 0.14, diameter = 8.2 $\mu m$), about 2 m in length. PSO was used to search a 15 dimensional space, each dimension corresponding to the voltage setting of each piezo actuator. PSO settings used were 30 particles and 80 iterations, with inertia damping ratio $w_{damp} = 0.99$, personal learning coefficient $c_{1} = 1.5$ and global learning coefficient $c_{2} = 2$. The cost function used was 2D correlation with respect to reference images generated from python module ofiber as seen in Figure \ref{fig:refs}. When fitting for $LP_{11}$ and $LP_{21}$, the reference images are rotated at various angles to find the best 2D correlation value (corr2).

\section*{Results}


\begin{figure}[hbt!]
\centering
\includegraphics[width=\linewidth]{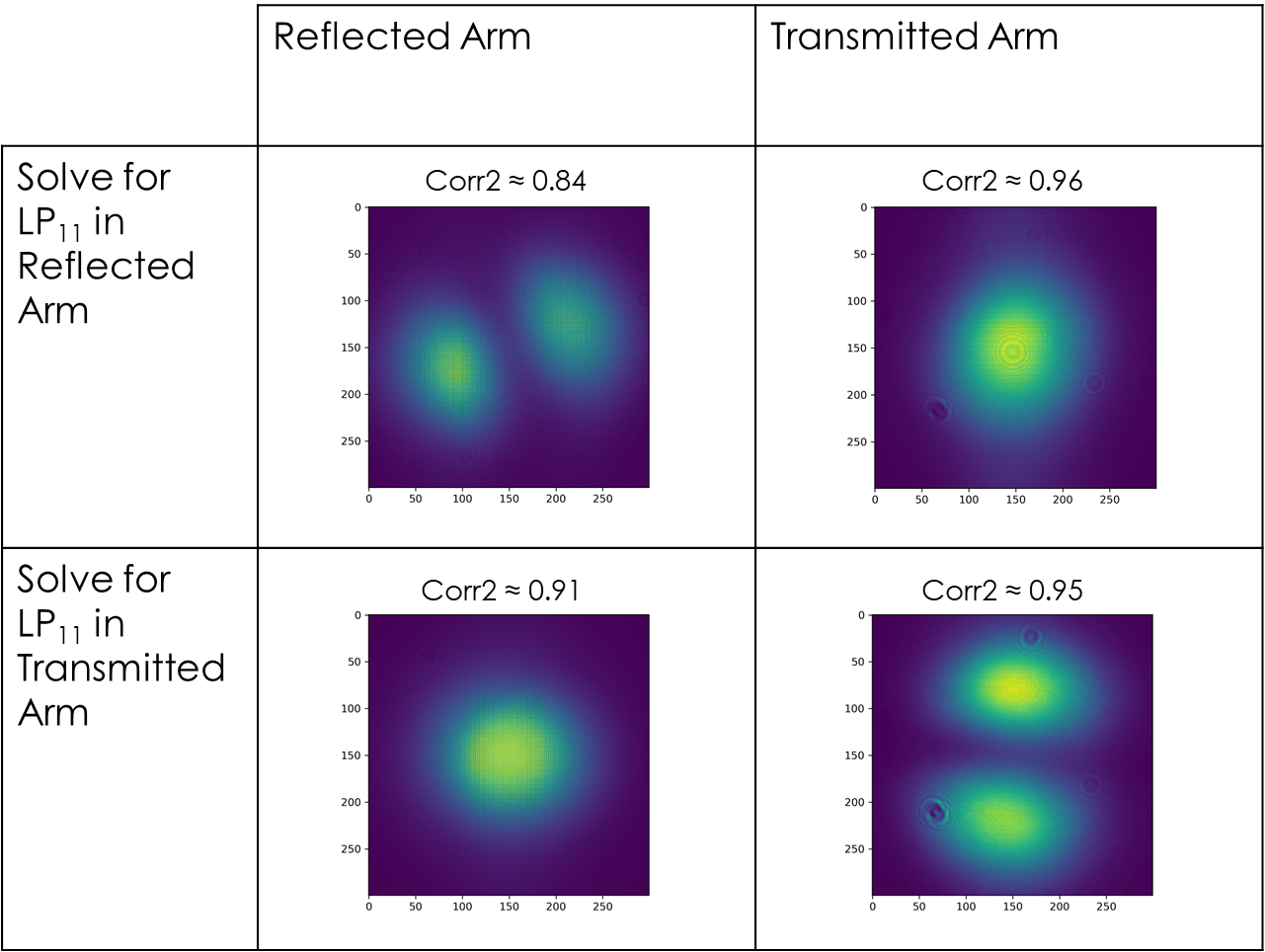}
\caption{Results obtained with the 808 nm laser when the PSO algorithm was set to solve for $LP_{11}$ in the reflected arm (top row) and transmitted arm of the PBS (bottom row). Corr2 values given are the 2D correlations of the images obtained with respect to $LP_{01}$ and $LP_{11}$ reference images. }
\label{fig:compiled808nm}
\end{figure}

\begin{figure}[hbt!]
\centering
\includegraphics[width=\linewidth]{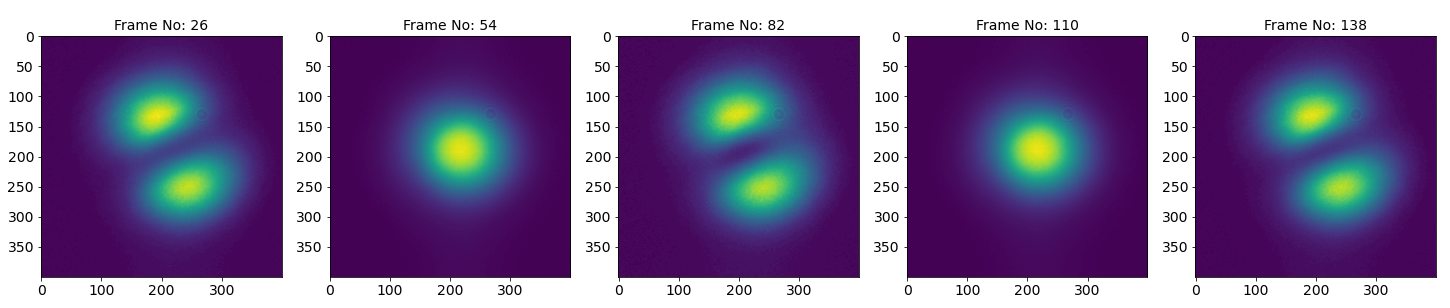}
\caption{Replication of fiber piano settings for $LP_{01}$ and $LP_{11}$ in SMF-28. The camera was running at about 27 frames per second, which makes each video frame displayed here about 1 second apart (See supplementary video). Swap rates can potentially enter the kHz range as the piezoelectric elements are rated for submillisecond response. }
\label{fig:swaptest}
\end{figure}

The setup used is in Figure \ref{fig:lasers} with SMF-28. With the 808 nm laser, The PSO algorithm was set to maximize the corr2 value for $LP_{11}$ in the reflected arm of the PBS, and then later in the transmitted arm. The results are shown in Figure \ref{fig:compiled808nm}. Interestingly, decent $LP_{01}$ images were obtained in the complementary arm even though the cost function did not take that into account. However, it should be noted that sub-optimal coupling efficiency was required for this result. When coupling efficiency was maximized, only $LP_{01}$ was observed, and the fibre piano was unable to produce $LP_{11}$ in any of the arms. 

Overall losses in generating the spatial modes were not severe. Input power entering the fibre piano was about 978 $\mu W$. When solving for $LP_{11}$ in the reflected arm, about 645 $\mu W$ was measured after the output reflective collimator, yielding a coupling efficiency of about 66$\%$. Of that 645 $\mu W$, about 333 $\mu W$ ($\approx 34\%$ of input power) was transmitted through the PBS and 251 $\mu W$ ($\approx 26\%$ of input power) was reflected. When solving for $LP_{11}$ in the transmitted arm, about 609 $\mu W$ was measured after the output reflective collimator, yielding a coupling efficiency of about 62$\%$. Of that 609 $\mu W$, about 153 $\mu W$ ($\approx 16\%$ of input power) was transmitted through the PBS and 390 $\mu W$ ($\approx 40\%$ of input power) was reflected. Thus the piano was able to perform mode separation with well tolerable losses.

Once the fibre piano has solved for a desired setting, the result could be consistently replicated at a later time so long as the setup had not been disturbed. This allows for the fibre piano to be used as a fast spatial mode switch, as seen in Figure \ref{fig:swaptest} and the supplementary video. The swapping speed is limited by the piezoelectric elements. ThorLabs states that the PB4NB2W models have sub-millisecond response time, so in principle a kHz swap rate is possible. 

\begin{figure}[hbt!]
\centering
\includegraphics[width=\linewidth]{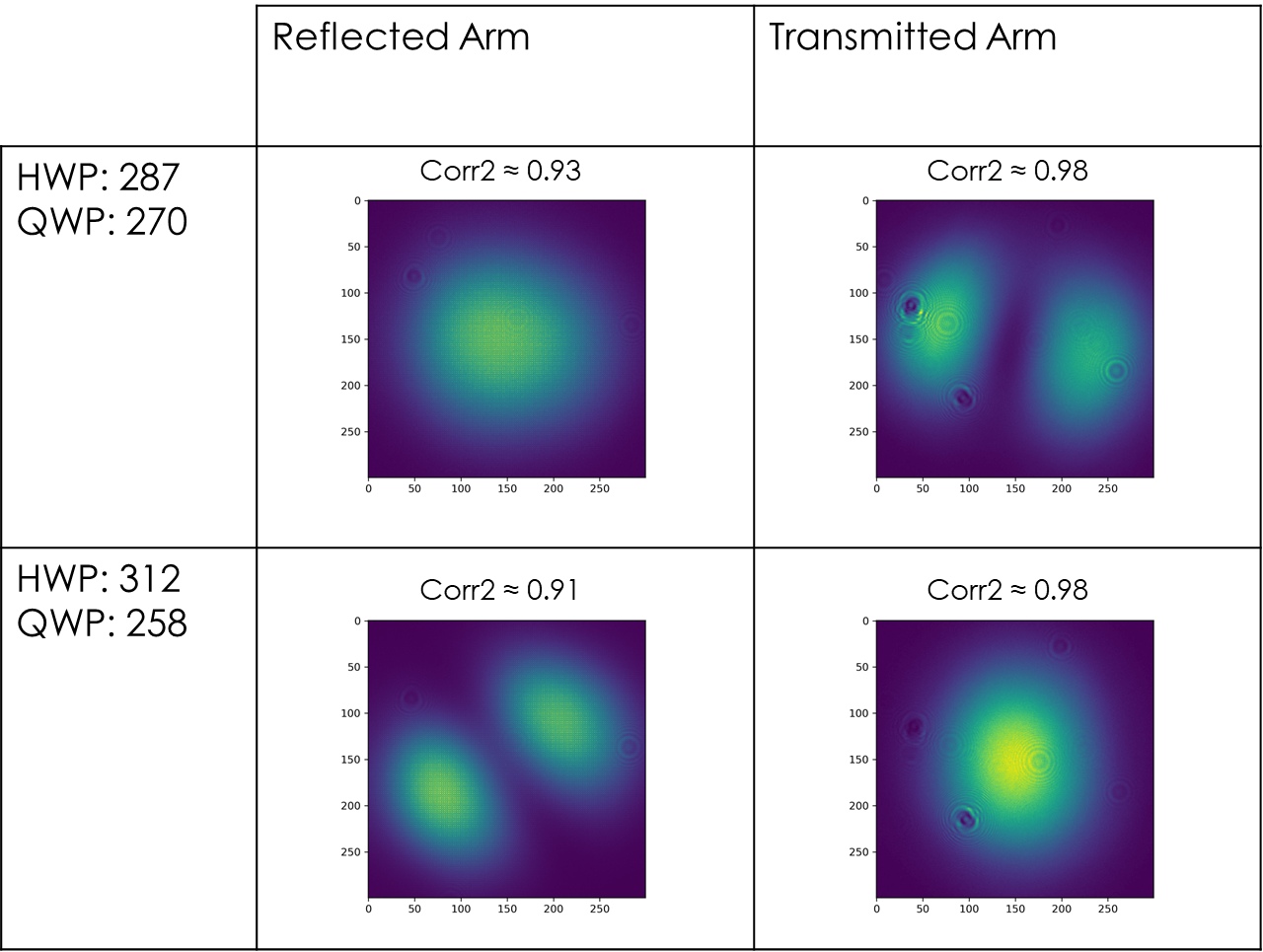}
\caption{The fiber piano was first used to solve for a configuration where $LP_{01}$ was in the transmitted arm and $LP_{11}$ in the reflected arm. The use of a HWP and QWP allowed for the modes to be swapped in both arms, albeit with $LP_{11}$ undergoing some rotation.}
\label{fig:waveplates}
\end{figure}

It was also observed that once the fibre piano had found a setting that produced $LP_{01}$ in the transmitted arm and $LP_{11}$ in the reflected arm, the modes in each arm could also be swapped by rotating the HWP and QWP in between the fixed Glan-Taylor prism and the coupling lens, as shown in Figure \ref{fig:waveplates}. The HWP produced more drastic changes while the QWP allowed for refinements in the mode shape. The input power into the fibre piano was 215 $\mu W$ and the output power right after the reflective collimator was 35 $\mu W$, a coupling efficiency of $16\%$. With the HWP at 312$^{\circ}$ and the QWP at 258$^{\circ}$, the transmitted $LP_{01}$ had 19.2 $\mu W$ ($\approx 9\%$ of input power) and the reflected $LP_{11}$ had 14.9 $\mu W$ ($\approx 7\%$ of input power). After rotating the HWP to 287$^{\circ}$ and QWP to 270$^{\circ}$, there was no significant change to coupling efficiency, but then the transmitted $LP_{11}$ had 6.1 $\mu W$ ($\approx 3\%$ of input power) and the reflected $LP_{01}$ had 28.8 $\mu W$ ($\approx 13\%$ of input power). It should be noted that without the use of the fiber piano, the waveplates themselves were insufficient to isolate desired modes given arbitrary initial conditions (ie coupling efficiency and fiber curvature).

\begin{figure}[hbt!]
\centering
\includegraphics[width=\linewidth]{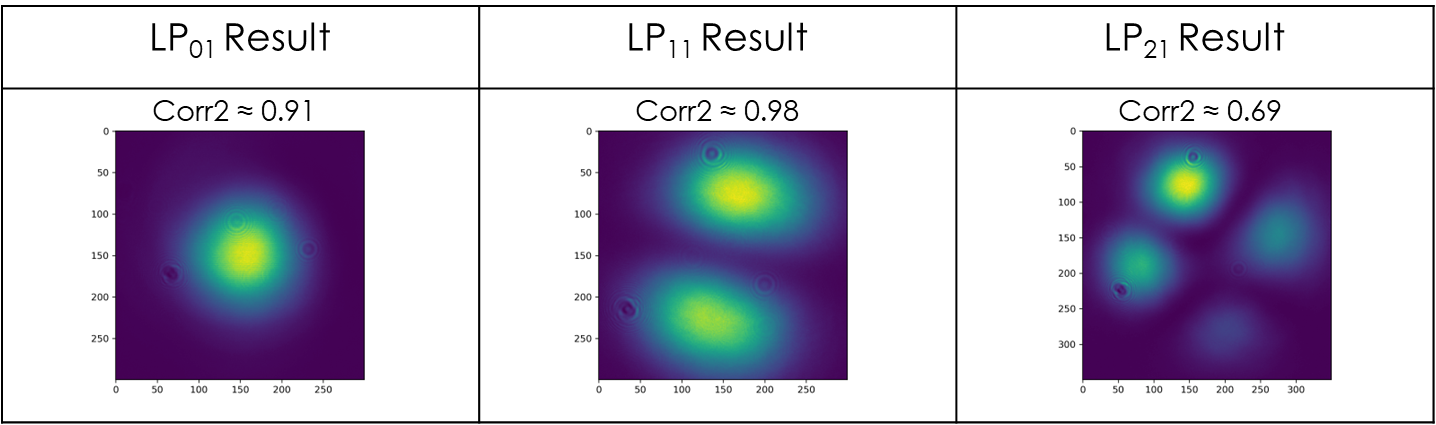}
\caption{Results obtained with the He-Ne (632.8 nm) laser when the PSO algorithm was set to solve for various spatial modes in the transmitted arm of the PBS. Obtaining $LP_{21}$ required a reduction in coupling efficiency of about $60\%$ compared to that for $LP_{01}$ and $LP_{11}$. Corr2 values given are the 2D correlations of the images obtained with respect to $LP_{01}$, $LP_{11}$ and $LP_{21}$ reference images. }
\label{fig:compiled633nm}
\end{figure}

Attempts were made to obtain higher order modes like $LP_{02}$ and $LP_{21}$ with the 808 nm laser to no avail, so a He-Ne laser (632.8 nm) was used instead in the hopes that the shorter wavelength allowed for those modes to be easily supported in the SMF-28. At the transmitted arm of the PBS, the fibre piano was able to produce $LP_{01}$, $LP_{11}$ and $LP_{21}$, but was unable to get $LP_{02}$ as shown in Figure \ref{fig:compiled633nm}. However, solving for $LP_{21}$ was only possible after reducing the power exiting the reflective collimator by about $60\%$ relative to that used to solve for $LP_{01}$ and $LP_{11}$. Even with that reduction in coupling efficiency, the fibre piano was still unable to get the 4 lobes with even shape and brightness. It appears that modes of higher order than $LP_{11}$ pose greater difficulty, but the potential to isolate them is definitely present.

\begin{figure}[hbt!]
\centering
\includegraphics[scale=0.55]{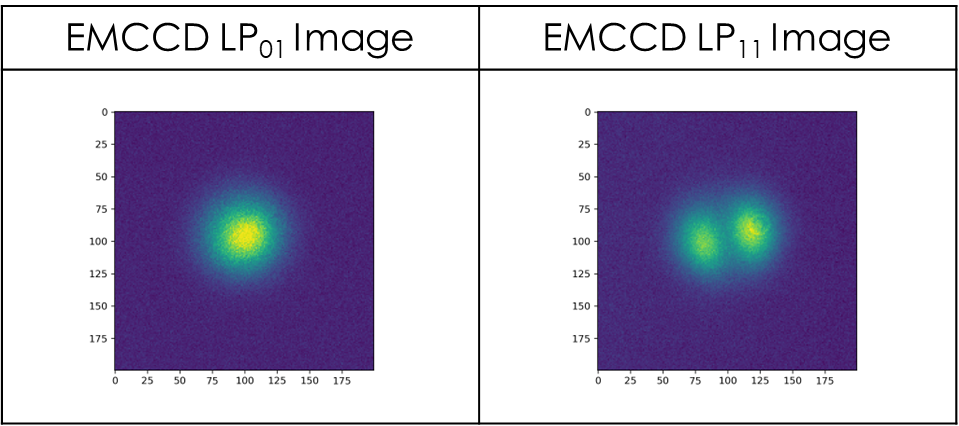}
\caption{$LP_{01}$ and $LP_{11}$ modes imaged by an EMCCD. This verifies that the fibre piano can also shape the spatial mode profiles of single photons.}
\label{fig:compiledEMCCD}
\end{figure}

Since the fibre piano could shape the spatial modes of laser light, it was reasonable to believe that it could do the same for single photons. To check this, the lasers were replaced with a degenerate photon pair source centred at 810 nm generated from Spontaneous Parametric Downconversion (SPDC) and an Electron-multiplying Charged Coupled Device (EMCCD) took the place of the camera in the reflected arm of the PBS. As the difference between the central wavelength of the photon pair source and that of the 808 nm laser was only a few nm, prior settings solved for by the piano using the 808 nm laser should be adequate for decent mode images. The results are shown in Figure \ref{fig:compiledEMCCD}. The distinct mode shapes are evident to the eye, confirming that the fibre piano is also able to shape the spatial modes of single photons.

\section*{Discussion}


The fiber piano displays strong potential as a spatial mode control tool, albeit with limitations that can be improved on. While unable to produce any desired spatial mode given arbitrary coupling conditions into the few mode fiber, this could be due to the ThorLabs PB4NB2W piezo actuators being only able to produce 450 $\mu$m of travel. It was found that when coupling efficiency was maximised in SMF-28 with either 808 nm or 632.8 nm lasers, only $LP_{01}$ was clearly excited and the fiber piano was unable to produce $LP_{11}$. Some degree of off-angle coupling was required for the fiber piano to access both modes. If stronger bends could be produced, stronger interactions between different spatial mode groups could be effected, affording the fiber piano better mode control and range of exploration. 

Further improvement in mode control could be achieved by incorporating the coupling angle as part of the PSO search space. Sharma and Chaudhuri have demonstrated mode control with this principle by using an electric field to deflect a fiber tip coated in an electrostrictive material \cite{Sharma2023}, thereby tuning the coupling angle. This would also allow for optimization of power transmission through the system while being able to access all desired modes at the output. In addition, the use of a HWP and QWP can provide additional tuning and even mode swapping at the outputs without requiring the fiber piano to search for a new configuration. Integrating them as part of the optimization process would also be simple with standard rotation mounts. 

However, it seems that the polarization and spatial mode degrees of freedom cannot be completely decoupled from each other. The PBS is essential for mode separation, and the fibre piano cannot work without it. In addition, input polarization greatly affects how modes are separated at the end. Observe that the $LP_{11}$ modes obtained at the transmitted and reflected arms are rotated relative to each other, and the piano was unable to generate $LP_{11}$ modes with a specific orientation. Also, the distribution of powers at both arms of the PBS are significantly influenced by input polarization. This could be due to the limitations in the piano's ability to transfer power between spatial modes of orthogonal polarizations, but more work will be needed to understand this. In any case, the transmission matrix of the optical fibre likely has fewer than $N^2$ controllable elements.    

Excitation and control of higher order modes beyond $LP_{11}$ is challenging but potentially possible. The main ways to access them more easily are increasing the core diameter of the fiber or using lasers of shorter wavelength. This work has demonstrated that a He-Ne laser is sufficient to obtain $LP_{21}$ in SMF-28, but with significantly higher coupling losses and lower corr2 values. Standard multimode fibres were tested as well but the fibre piano was unable to tame the resultant speckle patterns. That is an expected outcome, as the more spatial modes that can be excited in the fiber the more keys the fiber piano will require to isolate the desired modes \cite{Bromberg2020}. Custom made fibres with core sizes smaller than standard multimode fibres could be of use in going beyond $LP{21}$ while avoiding the speckled mess. However, coupling losses required to excite those modes are likely to be higher judging from the trend observed. An alternative could be modifying the spatial mode of the light entering the fibre piano. This work has used a Gaussian beam for the input, so the low overlap with higher order modes could be responsible for additional coupling losses being necessary. Launching light from a few or multimode fibre into the fibre piano should be sufficient to test this idea.  




\bibliography{sample}





\section*{Acknowledgements}

We thank M. Mukherjee (Centre for Quantum Technologies) for loaning his EMCCD.

\section*{Author contributions statement}

S.J.W and A.B. conducted the experiments and analysed the results. S.J.W. wrote the manuscript. A.S. designed the electronics used to control the piezoelectric elements. Z.F and R.S helped set up the fiber piano. A.L. supervised the study. Y.B. conceptualized and developed the fiber piano. All authors reviewed the manuscript. 

\section*{Availability of data and materials}
The data used in this study is available from the corresponding author on reasonable
request.

\section*{Additional information}

\subsection*{Competing interests}
The authors declare no competing interests.

\subsection*{Funding}
This research is supported by the National Research Foundation, Singapore (Grant NRF2020-NRF-ISF004-3538) and A*STAR under its CQT Bridging Grant.





\end{document}